\documentclass[conference]{IEEEtran}
\usepackage{cite}
\usepackage{amsmath,amssymb,amsfonts}
\usepackage{algorithmic}
\usepackage{graphicx}
\usepackage{textcomp}
\usepackage{xcolor}
\usepackage{url}
\usepackage[left=1.62cm,right=1.62cm,top=1.9cm]{geometry}

\begin{document}

\title{A BERT-based Empirical Study of Privacy Policies’ Compliance with GDPR
}

\author{Lu Zhang$^{1}$, Nabil Moukafih$^{1}$, Hamad Alamri$^{1}$, Gregory Epiphaniou$^{1}$, Carsten Maple$^{1}$\\$^{1}$Warwick Manufacturing Group, University of Warwick, Coventry, UK\\}

\maketitle

\begin{abstract}
Since its implementation in May 2018, the General Data Protection Regulation (GDPR) has prompted businesses to revisit and revise their data handling practices to ensure compliance. The privacy policy, which serves as the primary means of informing users about their privacy rights and the data practices of companies, has been significantly updated by numerous businesses post-GDPR implementation. However, many privacy policies remain packed with technical jargon, lengthy explanations, and vague descriptions of data practices and user rights. This makes it a challenging task for users and regulatory authorities to manually verify the GDPR compliance of these privacy policies. In this study, we aim to address the challenge of compliance analysis between GDPR (Article 13) and privacy policies for 5G networks. We manually collected privacy policies from almost 70 different 5G MNOs, and we utilized an automated BERT-based model for classification. We show that an encouraging 51$\%$ of companies demonstrate a strong adherence to GDPR. In addition, we present the first study that provides current empirical evidence on the readability of privacy policies for 5G network. we adopted readability analysis toolset that incorporates various established readability metrics. The findings empirically show that the readability of the majority of current privacy policies remains a significant challenge. Hence, 5G providers need 
to invest considerable effort into revising these documents to enhance both their utility and the overall user experience.

\end{abstract}


\section{Introduction}
\label{sec:intro}
Privacy laws are put in place to safeguard the personal information of individuals that is gathered by public or private organisations, governments, and other individuals. Among numerous data protection policies, the General Data Protection Regulation (GDPR) is regarded as one of the most stringent privacy rules. The main objective of GDPR is to empower individuals to have greater control over their personal information and to safeguard their rights concerning this data. In today's interconnected world, Cyber-Physical Systems (CPS) represent the nexus of computational and physical processes, functioning as intricate networks that interface with humans in sophisticated ways. These systems, characterized by their dynamic data exchanges and real-time responses, are dependent on reliable and swift communication infrastructures—hence the natural alignment with 5G networks. As CPSs become more integrated into daily life, they inevitably process personal and potentially sensitive data, whether it's location data from autonomous vehicles or health metrics from wearable medical devices. Such extensive data interaction accentuates the imperative of GDPR compliance. Ensuring that 5G MNOs—likely the communication backbone of many CPSs—are adhering to robust and clear privacy policies is not just about legal compliance. It's about fortifying trust in an ecosystem where digital and physical realms merge, and humans become intrinsic participants. If users or stakeholders struggle to understand how their data is being used due to jargon-laden policies, or if they suspect non-compliance, it could stymie the widespread adoption and evolution of CPS. Hence, studies focusing on enhancing the clarity and compliance of privacy policies in the 5G space are not merely topical—they are foundational to the successful integration of CPS into our societal fabric.

Since the implementation of GDPR in May 2018, businesses have been concentrating on transforming their data handling practices, which includes modifying privacy policies \cite{2018Privacy}. This is due to the fact that non-adherence to GDPR could result in penalties as high as 4$\%$ of a company's total yearly revenue or 20 million Euros \cite{GDPR}. Despite the severity of these fines, there are still numerous companies that have not yet achieved full compliance with GDPR. For instance, the UK's Information Commissioner's Office (ICO) imposed a 183 million Euro fine on British Airways \cite{BA} for not securing users' personal information and breaching GDPR. Since May 2019, the Irish Data Protection Commission (DPC) has initiated 19 investigations \cite{DPC} into potential privacy violations by large corporations (such as Google, Twitter, and Facebook).

However, it can be challenging to ascertain or verify if businesses or organizations that gather, handle, or retain individuals' private information are fully in compliance with GDPR. There are mainly two difficulties. 
First, GDPR is written in natural language, which incorporates numerous specialized terms that may be difficult for those lacking domain expertise to comprehend. In addition, privacy policies are typically long documents written in plain language, making it a time-consuming task for users to read. Due to these reasons, some researchers propose automatic compliance checking system. For example, the authors of \cite{hamdani2021combined} developed a combined rule-based and machine learning method to conduct automated compliance checking of privacy policies with GDPR. In \cite{torre2019using}, a generic model of the GDPR using UML class diagrams and object constraint language (OCL) constraints. The authors of \cite{amaral2023nlp} developed an automated solution that leverages natural language processing (NLP) technologies to check the compliance of a given data processing agreement under GDPR. In addition, the authors of \cite{aborujilah2022conceptual} proposed and developed a knowledge graph-based tool for GDPR contract compliance verification which binds GDPR’s legal basis to data sharing contracts.

In this paper, we investigate a BERT-based model to check whether the data controllers and processors in the 5G network are adhering to the GDPR's requirements for data protection and privacy. A 5G roaming scenario is considered to illustrate the compliance checking system in 5G network. When a user from the European Union (EU) travels to a non-EU country and uses their 5G mobile device to access the internet, their personal data may be transferred across borders. The GDPR requires that any transfer of personal data outside of the EU must be subject to appropriate safeguards. As a data controller, the mobile network operator is responsible for ensuring that they comply with the GDPR's requirements regarding the processing of personal data, including ensuring that appropriate measures are in place to protect against data breaches and unauthorized sharing with third parties. For example, the system can check if the mobile network operator is sharing the user's personal data with any third parties without their explicit consent. If such unauthorized sharing is detected, the system can alert the relevant authorities or take other appropriate actions to prevent the data breach from occurring. Therefore, implementing a compliance checking system for privacy policies can help the mobile network operator ensure that they are complying with these requirements and can help demonstrate their commitment to protecting their users' privacy rights. The BERT-based empirical study addresses several significant scientific and technical challenges. Firstly, it confronts the complexity of assessing GDPR compliance in privacy policies, especially given the pervasive technical jargon, protracted explanations, and ambiguous descriptions prevalent in many such documents. With the evolution of 5G networks and their distinct privacy implications, the study analyzes policies from a substantial sample of nearly 70 5G MNOs. To the best of our knowledge, this is the first work to check the compliance of privacy policies for 5G network. In addition, the study pioneers an empirical investigation into the readability of 5G privacy policies, emphasizing the urgent need for clearer documentation to benefit both users and regulatory authorities.

The rest of the sections are arranged as follows. Section II introduces the details of the GDPR Privacy Requirements. The experimental methodology is presented in Section III followed by the results and discussions in Section IV. Finally, conclusions are drawn in Section V.


\section{The GDPR Privacy Requirements}
\label{secII}
Since May 2018, the General Data Protection Regulation (GDPR) has been in effect with the purpose of safeguarding the privacy and data security of all European Union (EU) citizens. This regulation applies to any organization that processes personal information of EU residents, regardless of the organization's location. GDPR is Written in natural language, and it consists of 11 chapters and 99 articles.

According to \cite{gerl2019layered}, GDPR Articles 12-14 outline the rules regarding the type and manner of information that must be shared with data subjects. Article 12 emphasizes the methods for upholding the rights of these individuals, while Article 13 details the information that data controllers must supply when collecting personal data, making it appropriate for inclusion in privacy policies. Conversely, Article 14 specifies the required information when personal data has not been obtained.

Our study aims to examine potential GDPR violations by data controllers when collecting personal data. Hence, we concentrate on the provisions outlined in Article 13 of GDPR. According to \cite{liu2021have}, 10 different labels were extracted for Article 13 of GDPR. The details are given as below.
\begin{enumerate}
  \item Collect Personal Information: 
Gather information from data subjects that can be used to determine their personal IDs.
  \item Data Retention Period: 
Duration for retaining personal data.
  \item Data Processing Purposes: The purposes for handling personal information.
  \item Contact Details: Contact information for the data controller or Data Protection Officer.
  \item Right to Access: The data subject's right to request access to their personal information from the controller.
  \item Right to Rectify or Erase: The data subject's right to request the controller to correct or delete their personal information.
  \item Right to Restrict of Processing: The data subject's right to request the controller to limit processing related to the individual.
  \item Right to Object to Processing: The data subject's right to ask the controller to object to data processing.
  \item Right to Data Portability: The data subject's right to obtain and transfer their personal data to a different controller.
  \item Right to Lodge a Complaint: The data subject's right to file a complaint with a regulatory authority.
\end{enumerate}

\section{Methodology}
In this section, we present the methodology adopted in this work. The proposed method consists of two steps. In the first step, a Bert-based classification model is trained using the privacy policy dataset and the annotated GDPR requirements. In addition, an oversampling technique is utilized to improve the performance. In the second step, we assess the adherence of privacy policies to GDPR within the context of 5G networks.
\label{secIII}
\subsection{Training}
\subsubsection{Privacy policy collection and annotation}
The privacy policy and annotations used for training in this work is the same as the one in \cite{liu2021have}. Specifically, privacy policies of APPs were collected from Google Play, which is one of the most popular application stores. To gather privacy policies, the following guidelines are considered: (1) select policies from applications that rank highly on Google Play; and (2) ensure the policies come from various categories, as differing categories may necessitate distinct access to user information. The collected privacy policies span 22 application types, including Game, Communication, and Business. To maintain the quality of the collected documents, the filters are applied with the following criteria: (1) the privacy policy must be in English; (2) the content of the policy should not be too short, setting a lower limit of 2KB for the document size; and (3) duplicate privacy policies, often found in various apps from the same company, are eliminated. Initially, 1,313 privacy policies were gathered, and 304 valid policies remained for annotation after applying the filters. Those 304 privacy policies contains over 926K words and 36K sentences. The average word count for privacy policies in this corpus is 3,049, with 10$\%$ having fewer than 1,000 words and the shortest policy consisting of 154 words.

Then we explain the data annotation process. A total of 22 volunteers were enlisted, consisting of undergraduate and postgraduate students majoring in law and computer science, with strong English skills, to annotate the privacy policies. In order to maintain the quality of annotation, the volunteers were initially trained on the annotation task. A brief tutorial was provided, along with labeled example sentences to clarify the meaning of each label. After the training, the volunteers were asked to label a small set of privacy policies, and the quality of their annotations was assessed. Any misunderstandings were clarified as needed. Through this process, it was ensured that all volunteers had a clear understanding of the labels, allowing for control over annotation quality. Three volunteers were required to independently label each sentence. Sets of privacy policies were assigned to each volunteer, who completed his/her tasks individually. On average, annotating one privacy policy took 40 minutes for each volunteer. After all the tasks were completed, the three volunteers who labeled the same policies were asked to convene and consolidate their labels. In accordance with the standard process, if all three volunteers assigned the same label, it was used as the final label for the sentence. Otherwise, they discussed until a consensus was reached.

Table 1 presents the details of the annotated corpus, with the Frequency column indicating the cumulative count of each corresponding label within the corpus. Coverage represents the percentage of privacy policy documents featuring the respective label, while the Avg.W column signifies the average word count per sentence. Among the privacy policy labels, Data Processing Purposes (DPP) and Collect Personal Information (CPI) emerge as the most frequently occurring. Conversely, other categories such as Right to Lodge a Complaint (RLC), Right to Data Portability (RDP), Right to Restrict of Processing (RRP), and Right to Access (RA) are mentioned far less often. A unique label, "Other," encompasses all sentences not falling under these 10 labels. As the corpus aims to annotate entire privacy policy documents, all sentences are explicitly annotated, with the "Other" category accounting for 84$\%$ of the total sentences.

\begin{table*}[!ht]
\caption{The categorized Statistics on the annotated Corpus \cite{liu2021have}}
\begin{center}
\scalebox{1.1}{
\begin{tabular}{llll}
\hline Label                                 & Frequency & Coverage (\%) & Avg.W \\ \hline
\hline
Collect Personal information (CPI)    & 1,542     & 94.41          & 31.61 \\ 
Data Retention Period (DRP)           & 448       & 61.51          & 30.50 \\ 
Data Processing Purposes (DPP)        & 1,839     & 93.75          & 25.76 \\ 
Contact Details (CD)                  & 721       & 85.20          & 24.13 \\ 
Right to Access (RA)                  & 115       & 29.28          & 25.32 \\ 
Right to Rectify or Erase (RRE)       & 562       & 70.07          & 23.61 \\ 
Right to Restrict of Processing (RRP) & 127       & 29.28          & 23.03 \\ 
Right to Object to Processing (ROP)   & 245       & 40.46          & 23.24 \\ 
Right to Data Portability (RDP)       & 167       & 35.53          & 26.30 \\ 
Right to Lodge a Complaint (RLC)      & 145       & 36.84          & 24.77 \\ \hline
Other                                 & 30,669    & 100.00         & 24.98 \\ \hline
\end{tabular}}
\end{center}
\end{table*}

\subsubsection{Classification model}
In this paper, contextualized BERT model \cite{devlin2018bert} is adopted to perform the classification task. BERT \cite{devlin2018bert} is a prominent model for generating contextualized word embeddings, which has demonstrated exceptional results on numerous NLP tasks \cite{sun2019fine}. It takes an entire sentence as input and produces a series of hidden vectors using a highly pre-trained model. As suggested by Devlin et al. \cite{devlin2018bert}, we utilize the vector output from the sentence's initial symbol [CLS] in the final layer as a comprehensive sentence representation. Subsequently, we employ a feed-forward neural network (FFN) layer to assess the possibility of each potential label. This procedure can be summarized as follows:
\begin{equation}
\label{equ:Bert1}
s = \textup{BERT}(w_{1} \cdot \cdot \cdot \cdot w_{n})
\end{equation}
\begin{equation}
\label{equ:Bert2}
o = \textup{FFN}(s)
\end{equation}
We adhere to the standard process, proven to be successful across various tasks \cite{devlin2018bert, hao2019visualizing}, for fine-tuning the BERT parameters in conjunction with our task objective, and cross-entropy loss is used as the final objective function.

In addition, to balance the number of data for different classes in the training dataset, a random oversampling technique is adopted. Specifically, we randomly duplicate examples from the minority class in the training dataset. By doing this, the performance of the Bert-based classifier is further improved. The experimental results can be found in section IV.

\subsection{The compliance analysis process}
In this paper, we collected privacy policies for 5G Mobile network operators (MNOs) in both UK and EU countries. We collected privacy policies from 68 companies including O2, EE, Vodafone, Three etc. We use the trained BERT-based model to perform the compliance analysis for the privacy policies of 5G network. Specifically, with the labels extracted from Artical 13 of GDPR as shown in table I, 10 rules can be obtained including CPI, DPR, DPP, CD, RA, RRE, RRP, ROP, RDP, RLC. Each regulation specifies the type of information that the organization must provide to the data subject if an individual's information is collected. Given a privacy policy document, we need to check whether these 10 rules appear or not, in other words, this compliance analysis task is broken down into a sentence classification task, which involves examining the privacy policy to determine if it contains sentences labeled as required. If these 10 rules can be found in the privacy policy, then the GDPR regulation is not violated. Otherwise, a violation will be reported.

In addition, the readability of the privacy policy is also investigated. Readability served as an objective indicator of comprehensibility, offering an impartial numerical assessment. To evaluate the readability of privacy policies of 5G MNOs, a web-based readability calculator was employed to calculate and analyze the readability statistics. This web-based calculator was provided by WebpageFX, Inc, Harrisburg, PA \cite{Readability}. The Readability Test Tool was selected for its user-friendly interface. It should be noted that multiple free resources are available for calculating readability. The readability calculator provided various statistics, including word count, sentence count, Flesch reading ease (FRE), Flesh-Kincaid Grade Level (FKG), simplified measure of Gobbledygook (SMOG), and Automated Readability Index (ARI). To better illustrate the details of these readability metrics, we present a table as below.

\begin{table*}[!ht]
\caption{Overview of Readability Metrics}
\begin{center}
\scalebox{1.0}{
\begin{tabular}{|lll|}
\hline
\multicolumn{1}{|l|}{\textbf{Metric}} & \multicolumn{1}{l|}{\textbf{Formula}}                                                                             & \textbf{Score Mapping}                                                                                                                                                                                                                                                                                                               \\ \hline
\multicolumn{1}{|l|}{FRE}             & \multicolumn{1}{l|}{\begin{tabular}[c]{@{}l@{}}FRES = 206.835 -\\ 1.015 x ASL - 84.6 x\\ ASW\end{tabular}}        & \begin{tabular}[c]{@{}l@{}}90 - 100 = very easy = 4th grade; 80 - 90 = easy = 5th grade;\\ 70 - 80 = fairly easy = 6th grade; 60 - 70 = standard = 7th to\\ 8th grade; 50 - 60 = fairly difficult = some high school; 30 - 50\\ = difficult = high school or some college; 0 - 30 = very\\ difficult = college graduate\end{tabular} \\ \hline
\multicolumn{1}{|l|}{FKG}             & \multicolumn{1}{l|}{\begin{tabular}[c]{@{}l@{}}FKG = 0.39 x ASL +\\ 11.8 x ASW – 15.59\end{tabular}}              & US reading grade level                                                                                                                                                                                                                                                                                                               \\ \hline
\multicolumn{1}{|l|}{SMOG}            & \multicolumn{1}{l|}{\begin{tabular}[c]{@{}l@{}}SMOG = square root\\ of (SYW x 30 /\\ sentences) + 3\end{tabular}} & \begin{tabular}[c]{@{}l@{}}0 - 6 = low-literate; 7 = junior high school; 8 = junior high\\ school; 9 = some high school; 10 = some high school; 11 =\\ some high school; 12 = high school graduate; 13 - 15 = some\\ college; 16 = university degree; 17 - 18 = post-graduate\\ studies; 19+ = post-graduate degree\end{tabular}     \\ \hline
\multicolumn{1}{|l|}{ARI}             & \multicolumn{1}{l|}{\begin{tabular}[c]{@{}l@{}}ARI = 0.5 x ASL +\\ 4.71 x ALW - 21.43\end{tabular}}               & US reading grade level                                                                                                                                                                                                                                                                                                               \\ \hline
\multicolumn{3}{|l|}{\begin{tabular}[c]{@{}l@{}}Legend: ALW = Average number of letters per word; ASL = Average sentence length; ASW = Average number of syllables per word; LW =\\ Number of words with more than six characters;  SYW = Number of words with three or more syllables).\end{tabular}}                                                                                                                                                                                           \\ \hline
\end{tabular}}
\end{center}
\end{table*}


\section{Experimental results}
\label{secVI}
\subsection{Classification Results}
The classification results are shown in Table III. The optimum Precision, Recall, and F-score for each class are presented. The row called "Avg" displays the macro average across the 10 labels extracted from the GDPR. Notably, we provide the classification results for the "Other" category, which is not directly relevant to our compliance analysis task, but it comprises a significant quantity of sentences which impact the classification performance of the remaining 10 labels. Certain categories, such as CPI, present more challenges than others due to its inherent ambiguities. This largely stems from unclear definitions of personal information. Based on Article 4(1) of the GDPR, "personal data is any information which are related to an identified or identifiable natural person". Nevertheless, there are instances in privacy policies where the types of data collected are not clearly defined, or cases where non-personal data such as the user's browser version is collected, which does not align with the GDPR definition of personal information. Except category of CPI and RLC, we can obtain from the precision that our pre-trained Bert model performs well in terms of the classification, especially, for the category of RRP and RDP, the precision can both achieve 90 $\%$. Hence, due to its outstanding performance, we use pretrained Bert for the following compliance checking process. 
\begin{table}[]
\caption{Precision, Recall, and F-score for each class}
\begin{center}
\scalebox{1.1}{
\begin{tabular}{llll} 
\hline     & P    & R    & F    \\  \hline
\hline
CPI   & 0.77 & 0.89 & 0.83 \\
DRP   & 0.81 & 0.94 & 0.87 \\
DPP   & 0.81 & 0.84 & 0.83 \\
CD    & 0.85 & 0.90 & 0.87 \\
RA    & 0.86 & 0.82 & 0.84 \\
RRE   & 0.82 & 0.90 & 0.85 \\
RRP   & 0.90 & 0.85 & 0.87 \\
ROP   & 0.82 & 0.91 & 0.86 \\
RDP   & 0.90 & 0.94 & 0.92 \\
RLC   & 0.78 & 1.00 & 0.88 \\
Avg   & 0.83 & 0.89 & 0.86 \\\hline
Other & 0.96 & 0.93 & 0.94 \\ \hline
\end{tabular}}
\end{center}
\end{table}
\subsection{Compliance Analysis Results}
In this section, we perform compliance analysis for all the 68 5G companies. As illustrated in Figure 1, an encouraging 51$\%$ of companies demonstrate a strong adherence to GDPR, achieving an compliance rate between 80$\%$ and 100$\%$. On the contrary, a small but notable fraction, 12$\%$ of the companies, fall significantly short in their GDPR compliance, with their compliance level below 40$\%$.

Figure 2 presents an insightful analysis of the adherence to each specific GDPR rule. 91.18$\%$ of companies are found to comply with the CPI rule, implying that the majority of companies are engaging in personal information collection. Moreover, over 80$\%$ of companies' privacy policies can address the DPI, DPP, and CD guidelines. However, it's noteworthy that adherence to the rules of RRP and RDP is considerably lower, with only 47.06$\%$ and 57.35$\%$ of companies in compliance respectively. Across all the 5G companies under consideration, the average level of GDPR compliance reaches 72.94$\%$. To verify the experimental results of our model, we manually verified the compliance of a few random companies' privacy policies. We found that the results are accurate and in accordance with those obtained by our model.

\setlength{\textfloatsep}{0.6mm}
\begin{figure}[ht]
\centering
\includegraphics[scale = 0.6]{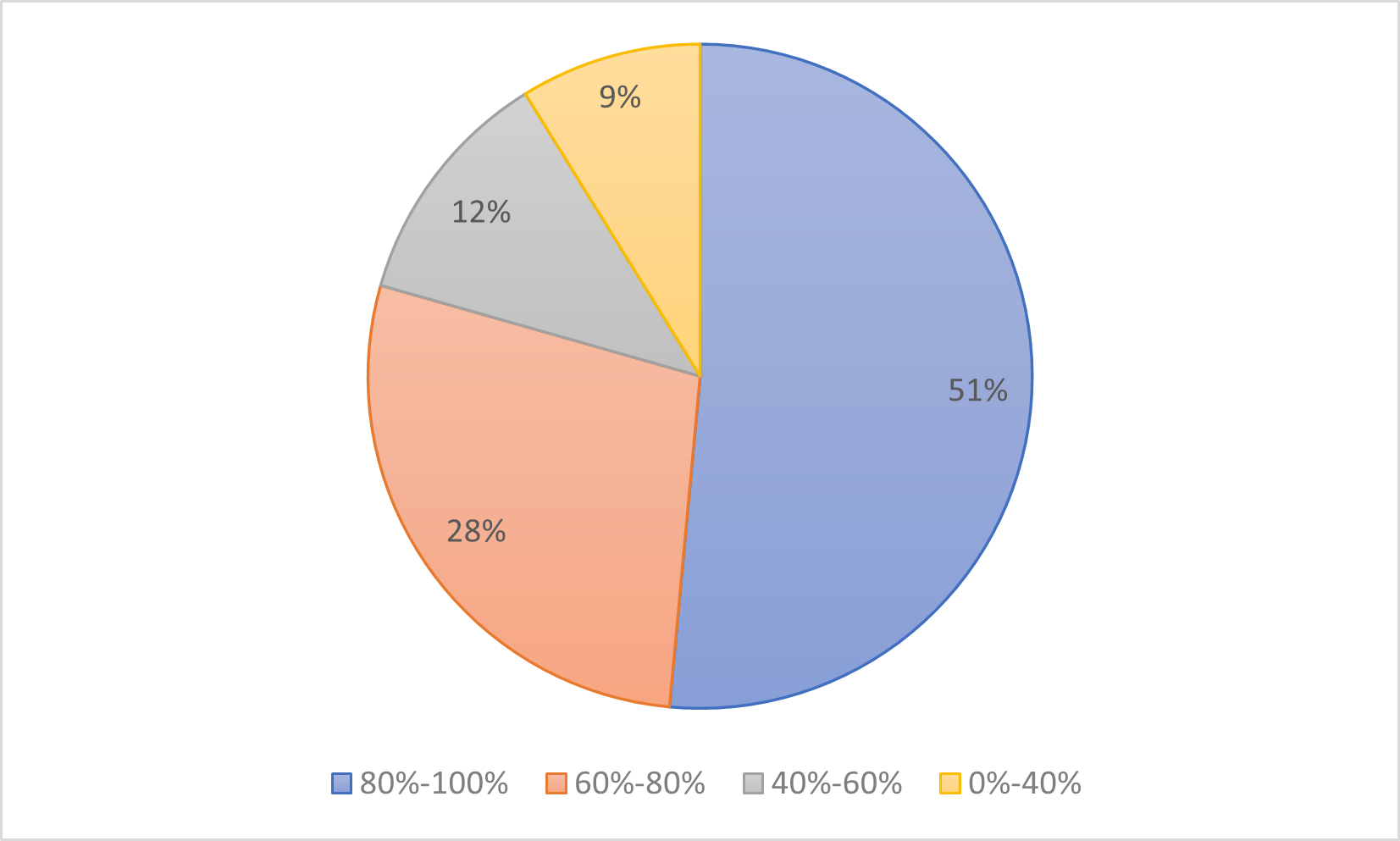}
  \caption{Percentage of the compliance for the 5G companies.}
  \label{fig:compliance distribution}
\end{figure}

\setlength{\textfloatsep}{0.6mm}
\begin{figure}[ht]
\centering
\includegraphics[scale = 0.6]{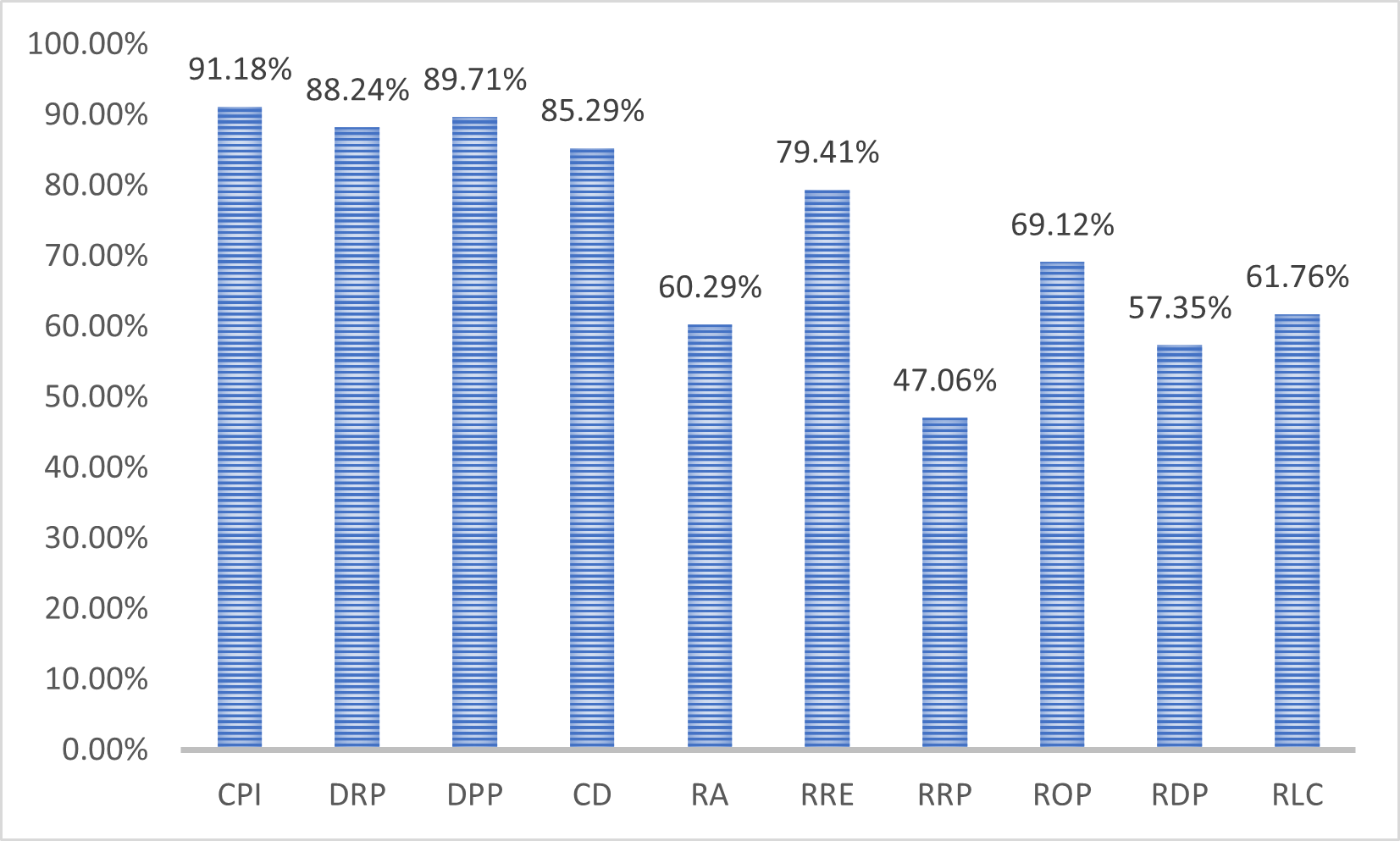}
  \caption{Percentage of the compliance for each rule among all the 5G companies.}
  \label{fig:each label}
\end{figure}

\subsection{Readability Analysis Results}
For each policy, the number of words, sentences and the number of words in one sentence have been calculated. In addition, the readability measures have also been investigated.

The privacy policies show a wide variation in length, with the shortest being 354 words and the longest exceeding 13,628 words. Similarly, the sentence count varies from a mere 17 to well over 604 in a single policy. On a typical basis, these policies tend to consist of around 4085 words and approximately 155 sentences. The word count per sentence in the privacy policy varies significantly, from a minimum of 13.5 to a maximum of 75.1 words. However, on average, a sentence in these policies contains about 28 words.

Based on the FRES, the comprehensibility of privacy policies for 5G MNOs varies significantly, ranging from 'standard' (with the highest FRES at 69.66) to 'very difficult' (lowest FRES at 10.32). This suggests that, on average, these policies tend to be quite difficult to understand. 

Based on the SMOG assessment, the readability of privacy policies, considering all the investigated domains, requires an educational background that spans from low-literacy to postgraduate degree levels for easy comprehension. It suggests that, on average, the required educational background to fully understand these policies is a postgraduate degree.

FKG,  ARI suggest the average US reading grade level of
14.05 (sd = 3.0),  14.29 (sd = 3.68) study years required by readers to comprehend privacy policies for the 5G network, respectively.

\section{Conclusion}
In our study, we delve into the degree of adherence and effectiveness of privacy policies with respect to GDPR. We show only a small fraction, 12$\%$ of the companies, fall significantly short in their GDPR compliance, with their compliance level below 40$\%$. In addition, we undertook an in-depth analysis of these documents using various readability evaluation tools. Our results highlight the necessity for 5G service providers to devote significant resources to refining these documents, with the aim of improving their readability and enriching the overall user experience.
\label{secV}

\bibliographystyle{IEEEtran}
\bibliography{refs}

\end{document}